# Spectrum standardization for laser-induced breakdown spectroscopy measurements

Zhe Wang, Lizhi Li, Logan West, Zheng Li*, Weidou Ni

*State Key Lab of Power Systems, Department of Thermal Engineering, Tsinghua-BP Clean Energy Center, Tsinghua University, Beijing, 100084, China*

**Abstract:** This paper presents a spectra normalization method for laser-induced breakdown spectroscopy (LIBS) measurements by converting the recorded characteristic line intensity at varying conditions to the intensity under a standard condition with standard plasma temperature, degree of ionization, and total number density of the interested species to reduce the measurement uncertainty. The characteristic line intensities of the interested species are first converted to the intensity at a fixed temperature and standard degree of ionization but varying total number density for each laser pulse analysis. Under this state, if the influence of the variation of plasma morphology is neglected, the sum of multiple spectral line intensities for the measured element can be regarded proportional to the total number density of the specific element, and the fluctuation of the total number density, or the variation of ablation mass, was compensated for by the application of this relationship. In the experiments with 29 brass alloy samples, the application of this method to determine Cu concentration shows a significant improvement over generally applied normalization method for measurement precision and accuracy. The average RSD value, average value of the error bar, $R^2$, RMSEP, and average value of the maximum relative error were: 5.29%, 0.68%, 0.98, 2.72%, 16.97%, respectively, while the above parameter values for normalization with the whole spectrum area were: 8.61%, 1.37%, 0.95, 3.28%, 29.19%, respectively.

**Keywords:**   LIBS; Normalization; quantitative measurement; plasma property

# 1 Introduction

For laser-induced breakdown spectroscopy (LIBS) measurement, a laser pulse focused by a lens acts as the excitation source, striking the sample surface to form a plasma with high temperature and high electron density, and a spectrometer then records the plasma emission spectrum to analyze its elemental composition and elemental concentrations constituting the samples. This technology has the following advantages compared with conventional spectroscopic analytical techniques [1, 2]: 1) LIBS is suitable for almost all sample phases (solid, liquid, gas); 2) LIBS has a fast response speed, suitable for real-time measurement; 3) LIBS requires little or no sample preparation; 4) there is little or no damage to the sample during analysis; and 5) LIBS has the capability of measuring most the chemical elements. Because of the above advantages, LIBS is widely applied across many fields. However, in LIBS measurement for one uniform sample, even under the same experimental parameters setup, plasma characteristics will vacillate due to fluctuations in laser energy and laser-sample interaction. This leads to larger signal uncertainties in the LIBS measurement compared with other atomic emission spectroscopy such as inductive coupled plasma (ICP). The drawback is a major impediment to both LIBS measurement precision and accuracy improvement and the commercialization development of LIBS [3]. Reducing LIBS signal uncertainties has been an important research focus.

One common method to reduce LIBS signal uncertainties is through repeated measurements and averaging of the LIBS spectra [4-8]. However, because of the non-linear relationship between spectral line intensity and plasma temperature and electron density, the linear averaging method has some limitations. Moreover, in certain cases such as online rapid chemical analysis and/or heterogeneous sample analysis, averaging of multiple measurements to reduce uncertainty is not feasible, and then LIBS is performed using a single laser shot [3]. Methods that reduce the uncertainty of each single pulse measurement are of necessity.

In data processing, one of the common methods for reducing the uncertainty is normalizing the data with the whole spectrum area [9-12], which can partially reduce the signal uncertainty resulted from fluctuations of laser energy, delay time, and gate time etc. [10]. The internal calibration method [13-17] is another approach, which can largely compensate for signal fluctuations. However, for internal calibration method, the spectral lines of the element under analysis and the internal calibration element should have the same or at least similar excitation energy of the upper level to offset the intensity ratio dependence on temperature, and the internal calibration element should have a fixed mass concentration in samples. The continuous background spectrum [18-22] or other signals generated by the plasma, such as acoustic waves [23, 24], can also be used to reduce the uncertainty. Overall, these methods, taking advantage of the correlation between laser energy, the plasma ablation mass, and the LIBS signal, can indirectly and partly reduce the LIBS signal fluctuations. A potentially better method is to use plasma characteristic parameters to directly reduce the spectral signal fluctuations.

In a previous paper [25], plasma temperature and electron number density (degree of ionization) were utilized to normalize the spectrum and they directly compensate for signal fluctuations caused by these two plasma characteristic parameters. However, the remaining uncertainty, resulting from variations in the total number density of particles in the plasma and plasma shape, was totally not compensated for. So the final signal still had a large uncertainty.

This article try to further compensate for the fluctuation of the line intensity due to the fluctuation of total number density of the measured element particles to reduce the uncertainty and, combined with the multi-pulse averaging method, improve the LIBS measurement accuracy. This method, in essence, transfer the selected characteristic line intensity at real application condition to a standard status with fixed plasma temperature, electron density, and ablation mass for different samples, therefore, we call it "spectrum standardization" instead of normalization to show the difference and improvement.

## 2 Theory

As the exact number of total number density of the measured element particles in plasma is almost impossible to determine without advanced auxiliary facility, directly converting the recorded spectral line intensity to the intensity under a standard total number density is not realistic. In this paper, we indirectly convert the spectral line intensity under varying total number density of the measured element to the intensity under a standard total number density by taking the below assumptions. After converting the characteristic spectral line intensities of specific element to the intensity under a standard plasma temperature and standard ratio between the ion and atom number density of the element of interest, the spectral line intensity is proportional to the total number density of the specific element particles neglecting the inference effects due to plasma morphology variation for the same sample. Using this relationship, the proposed approach compensates for the total particle number density fluctuations of multiple measurements and reduces the measurement uncertainty significantly. The specific principle and detailed process of the proposed method are described below.

### 2.1 Plasma temperature and degree of ionization standardizations

Under local thermodynamic equilibrium (LTE) conditions, atomic characteristic spectral line intensity in the LIBS measurement can be calculated by [26],

$$I_{ij}^I = F n_s \frac{n^I}{n^{II} + n^I} \frac{g_i \exp(-E_i/(kT))}{U^I(T)} A_{ij} \qquad (1)$$

where superscript $I$ and $II$ represent the measured-element atom and ion, respectively, and the subscript $i$ and $j$ are the upper energy level and lower energy level for the characteristic line. $I$ is the characteristic spectral line intensity. $F$ is a constant instrument parameter for the determinate experimental condition. $A$, $g$ and $U(T)$ are the transition probability, statistical weight and partition function at temperature $T$, respectively. $E$ and $k$ are the excited energy and the Boltzmann constant. $n_s$ is the total number density of specific element species in the plasma. $n^I$ and $n^{II}$ are the number densities of the neutral and the singly ionized species of the same element, respectively. For typical LIBS measurements, $n_s = n^I + n^{II}$. The degree of ionization, $r$, defined here as the ratio between the singly ionized and neutral species of the same element, can be calculated by [27],

$$r = r(T, n_e) = \frac{n^{II}}{n^I} = (\frac{2\pi m_e kT}{h^2})^{3/2} \frac{2}{n_e} \frac{U^{II}(T)}{U^I(T)} \exp(-\frac{E_{ion} - \Delta E}{kT}) \qquad (2)$$

where $n_e$ is the electron number density in plasma, $h$ is Planck's constant, $E_{ion}$ is the ionization potential of the neutral species in its ground state, and $\Delta E$ is the ionization potential lowering factor with a typical value on the order of 0.1 eV.

The ion characteristic spectral lines intensity can be calculated by,

$$I_{mn}^{II} = F n_s \frac{n^{II}}{n^{II} + n^I} \frac{g_m \exp(-E_m / (kT))}{U^{II}(T)} A_{mn} \qquad (3)$$

Much literature already exists on the solutions for plasma temperature, electron density, and degree of ionization [25, 28-33]. In this paper, the plasma temperature and electron number density are obtained through the Boltzmann plot and Starks broadening method. Using the following formulas, the measured element characteristic spectral line intensity can be converted to the intensity under standard plasma temperature and standard degree of ionization. In this paper, a value close to the average values of temperatures and degree of ionization for all multiple measurements of multiple samples were taken as the standard value of plasma temperature and the standard degree of ionization. Standardization formulas for atomic and ionic spectral lines are given by, respectively

$$I(T_0, r_0) = I_{ij}^I \frac{r+1}{r_0 + 1} \frac{U^I(T) \exp(-E_i / kT_0)}{U^I(T_0) \exp(-E_i / kT)} \qquad (4)$$

$$I(T_0, r_0) = I_{mn}^{II} \frac{r_0(r+1)}{r(r_0 + 1)} \frac{U^{II}(T) \exp(-E_m / kT_0)}{U^{II}(T_0) \exp(-E_m / kT)} \qquad (5)$$

where $T_0$ is the standard plasma temperature, $r_0$ is the standard degree of ionization, and $I(T_0, r_0)$ is the calculated characteristic line intensity at a status with standard plasma temperature $T_0$ and standard degree of ionization $r_0$.

## 2.2 Total number density of the specific element standardization

After carrying out the steps in Section 2.1, the characteristic spectral line intensities had been converted to conditions with standard temperature and standard degree of ionization. Ideally, the remaining uncertainty of the characteristic spectral line intensity should come from the fluctuation of ablation mass in the single measurement. That is, ignoring the plasma space morphology variation, the intensity fluctuation should then be proportional to variation in the total number density of the specific element particles in the plasma for the same sample. That is,

$$I(T_0, r_0) - I(n_{s0}, T_0, r_0) = k_1(n_s - n_{s0}) \tag{6}$$

where, $n_s$ is the total number density of the specific element particles in plasma for single LIBS measurement, $n_{s0}$ is the total number density of the specific element particles in the standard plasma state with the same temperature, degree of ionization, and ablation mass for every single measurement, and $I(n_{s0}, T_0, r_0)$ is the spectral line intensity under standard plasma condition with standard total number density of the specific elemental species, standard plasma temperature, and standard degree of ionization. If the self-absorption and inter-element interference effects on the measurement spectrum can be neglected, the coefficient, $k_1$, should not change too much for different samples with the measurement species over a wide range. That is, this equation can be applicable for a sample set with similar matrix.

Under ideal conditions, the standard characteristic spectral line intensity $I(n_{s0}, T_0, r_0)$ is no longer affected by the measurement uncertainty. If ignoring the self-absorption and inter-element interference effects, the characteristic spectral line intensity is determined only by the element mass concentration, that is,

$$I(n_{s0}, T_0, r_0) = k_2 C + b \tag{7}$$

Also, we argue that, after normalization with plasma temperature and degree of ionization, the total number density of the specific element particles, $n_s$, should be proportional to the sum intensity of multiple characteristic spectral lines $I_T(T_0, r_0)$, while the standard total number density of the specific element particles $n_{s0}$ is proportional to the mass concentration of the sample. So,

$$n_s = k_3 I_T(T_0, r_0) \tag{8}$$

$$n_{s0} = k_4 C \tag{9}$$

In addition, ideally the intensity of each characteristic spectral line of the specific element should be proportional to the total number density of the specific element particles. That is, each characteristic spectral line of specific element can be used to compensate for the fluctuation from the total number density of the specific element particles. However, because a single characteristic spectral line is vulnerable to inter-element interference and other matrix effect, utilizing only one characteristic line intensity to represent the ablation mass fluctuations may entail large errors in the practical application. Besides, the spectral line that applied for concentration calculation cannot be utilized for uncertainty reduction since its function of compensating for the fluctuation will be overwhelmed by the concentration indication during the best curve fitting process in obtaining the model coefficients. In theory, normalization by the whole spectrum area after normalization using plasma temperature and degree of ionization for all spectral lines should also have the same functionality as the proposal approach below, but because the process requires a prohibitively large amount of calculation, it is not recommended. Therefore, this approach utilizes the sum intensities of multiple characteristic spectral lines of specific elements to represent the total number density of the specific element species as shown by Eq.8. Using a combination of Eqs.6, 7, 8, and 9, we obtain,

$$I(T_0, r_0) = a_1 C + a_2 I_T(T_0, r_0) + b \tag{10}$$

where $a_1 = k_2 - k_1 k_4$, $a_2 = k_1 k_3$.

Then, by substitution of Eq. 10 in Eq. 7, we obtain,

$$I(n_{s0}, T_0, r_0) = A_1 I(T_0, r_0) + A_2 I_T(T_0, r_0) + A_3 \tag{11}$$

where $A_1 = k_2 / a_1$, $A_2 = -k_2 a_2 / a_1$, $A_3 = -k_2 b / a_1 + b$.

Eq. 11 shows the way to standardize the characteristic spectral line intensity to the intensity under standard plasma temperature, standard degree of ionization, and standard total number density of measurement element species, while the Eq. 10 represents the spectra standardization model for samples' elemental concentration prediction.

In comparison with the model of the previous article [25], this model adds a new variable $A_2 I_T(T_0, r_0)$, which is taken to compensate for the fluctuation of the total number density of the specific element species. The essence of the model can be more clearly seen by rewriting Eq. 10 as

$$I(n_{s0}, T_0, r_0) = I(T_0, r_0) + k_1(k_4 C - k_3 I_T(T_0, r_0)) = k_2 C + b. \tag{12}$$

As seen, the sum of multi-line intensity is applied to correct the line intensity with standard temperature and degree of ionization from varying total number density to standard total number density. This also indicated that any other signal that contained in the full spectra with closely correlation with the total number density variation can be applicable for the compensation.

Basically, as laser pulse energy increase, the ablation mass and plasma increase as well as the whole spectral area. Therefore, normalization with whole spectrum area can be applied to compensate for the variation in laser energy. However, with the same laser energy, more ablation mass means lower plasma temperature, this means whole spectrum area normalization cannot compensate for the fluctuations due to laser-sample interaction variation. Moreover, as temperature increases, normally the degree of ionization will also increase, while these two changes have inversed impact on the change of the whole spectrum area, further making the whole spectrum area normalization less effective in reducing the uncertainty due to the combination variation in total ablation mass, temperature, and degree of ionization. The standardization approach overcomes the shortage of the whole spectrum area normalization method by reducing the impact of these three factors one by one directly, therefore combining the advantages of both the direct normalization method with plasma temperature and electron number density proposed in the previous paper [25] and the generally applied normalization with whole spectrum area method. Besides, it can be combined with the multi-pulse averaging method to not only reduce the uncertainty of every single measurement, but also the precision and accuracy of the multi-pulse analyses.

## 3 Experiment setup

This experiment utilized the Spectrolaser 4000 (XRF, Australia), which is described in detail in a previous paper [25].

To increase reliability of the model, a total of 29 brass alloy samples were used in the experiment, including a ZBY series of samples coming from the Central Iron and Steel Research Institute (CISRI) of China and brass alloy samples from the Shenyang Nonferrous Metals Processing Factory. The mass concentrations of the major elements (Cu, Zn, Pb, Fe) in the samples are listed in Table1.

Table 1.Major elemental concentrations of the brass alloy samples

| No. of sample | Sample serial No. | Cu（%） | Zn（%） | Pb（%） | Fe（%） |
|---|---|---|---|---|---|
| 1 | ZBY901 | 73 | 23.99 | 2.77 | 0.028 |
| 2 | ZBY902 | 60.28 | 38.79 | 0.766 | 0.047 |
| 3 | ZBY903 | 64.43 | 33.45 | 1.87 | 0.036 |
| 4 | ZBY904 | 59.14 | 38.85 | 1.5 | 0.167 |
| 5 | ZBY905 | 58.07 | 39.59 | 1.81 | 0.11 |
| 6 | ZBY906 | 56.62 | 41.76 | 0.581 | 0.037 |
| 7 | ZBY907 | 59.55 | 34.92 | 3.06 | 0.502 |
| 8 | ZBY921 | 59.89 | 39.01 | 0.318 | 0.288 |
| 9 | ZBY922 | 61.88 | 37.53 | 0.108 | 0.116 |
| 10 | ZBY923 | 69.08 | 30.44 | 0.018 | 0.052 |
| 11 | ZBY924 | 80.9 | 18.75 | 0.017 | 0.11 |
| 12 | ZBY925 | 85.06 | 14.79 | 0.029 | 0.028 |
| 13 | ZBY926 | 90.02 | 9.76 | 0.0084 | 0.024 |
| 14 | ZBY927 | 95.9 | 4.02 | 0.0028 | 0.012 |
| 15 | 1 | 96.86 | 3.06 | 0.0082 | 0.024 |
| 16 | 2 | 95.1 | 4.78 | 0.0236 | 0.066 |
| 17 | 3 | 94.46 | 5.26 | 0.05 | 0.182 |
| 18 | 4 | 92.7 | 6.81 | 0.098 | 0.336 |
| 19 | 5 | 89.97 | 9.83 | 0.0301 | 0.124 |
| 20 | 6 | 90.76 | 9.15 | 0.012 | 0.051 |

| 21 | 7  | 85.49 | 14.41 | 0.0283 | 0.097 |
| 22 | 8  | 79.1  | 20.74 | 0.029  | 0.098 |
| 23 | 9  | 70.44 | 29.04 | 0.132  | 0.182 |
| 24 | 10 | 69.25 | 30.66 | 0.0105 | 0.016 |
| 25 | 11 | 67.59 | 32.17 | 0.06   | 0.101 |
| 26 | 13 | 64.32 | 35.51 | 0.0697 | 0.067 |
| 27 | 14 | 63.42 | 36.18 | 0.163  | 0.14  |
| 28 | 15 | 60.81 | 38.59 | 0.294  | 0.236 |
| 29 | 16 | 57.98 | 41.04 | 0.591  | 0.427 |

To reduce the experimental error, test equipment was preheated for more than half an hour before the experiment, improving equipment stability. In addition, the brass alloy sample surfaces were scrubbed using lens-cleaning paper moistened with analytically pure anhydrous ethanol and air-dried to ensure the sample surface uniformity and cleanliness. In order to thoroughly eliminate experimental error due to sample surface contamination, a 150mJ laser beam was utilized to clean sample surface.

After optimization of the experimental parameters, the analysis-laser energy and delay time were set for 90mJ/pulse and 2.25μs, respectively. Under these setup parameters, the spectral signal-to-noise ratio was larger but the spectral signal intensity did not exceed the saturation spectral line intensity to the spectrometers.

Next, each sample was placed in the sample chamber in ambient air and atmospheric pressure first. To begin the test, the 150 mJ laser fired a single beam on the sample surface for cleaning before each measurement. For each sample, multiple measurements were taken at different locations on the sample. For measurements Nos.1 to 35, the analysis laser (90mJ/pulse) was fired at the sample, and a spectrometer was utilized to collect the spectra. For the 36th measurement point on the sample, background noise spectrum was recorded by the spectrometer using a laser pulse with much lower energy (10 mJ) followed by a sufficiently long delay time (19μ s).Thus, for each sample, there were 35 measured spectra and a background noise spectrum recorded. For the final background noise spectrum, an average of the background noise spectra for each of the 29 samples was taken.

With regard to pre-processing of the experimental data, the background noise spectrum was first subtracted in order to eliminate the interference signal form the external environment and the apparatus, limiting their influence on the experimental results. Then, each spectrum was corrected with white light, thereby eliminating spectral line intensity distortion resulting from the optical system sensitivity different for different spectral line wavelengths.

## 4 Results and discussion

In this section, the traditional uni-variate models without normalization, normalization with whole spectrum area, and with the newly proposed standardization method are compared.

To evaluate the quality of various models and normalization methods, five performance indexes were used as evaluative parameters: the relative standard deviation (RSD) of the spectral line intensity, the standard error (error bar) of the model predicted mass concentration, the coefficient of determination ($R^2$) of calibration curve, the root mean square error of prediction (RMSEP) of the mass concentration, and the maximum relative error of the model predicted mass concentration. RSD was applied to assess fluctuations of spectral line intensity. Error bar was utilized to value the precision of the model predicted mass concentration. $R^2$ was utilized for quality evaluation of calibration model. RMSEP and the maximum relative error of the model predicted mass concentration were used to measure the accuracy of model predictions. The smaller the RSD of spectral line intensity and the smaller the standard error of prediction mass concentrations, the more precise the LIBS measurement. The closer the $R^2$ value is to 1, the RMSEP closer to 0, and the maximum relative error closer to 0, the more accurate the LIBS measurement.

The 29 samples were arranged in accordance with the Cu concentration from smallest to largest. One of every three samples was selected as an evaluation sample to ensure that the evaluation samples represented an equal distribution across the full range of sample concentrations and best integrate the predictive capabilities of the model. According to this principle, a total of 9 samples (ZBY903, ZBY905, ZBY921, ZBY922, 2, 6, 7, 8, and 10) were selected as validation samples. The remaining 20 samples were used for calibration.

## 4.1 Baseline

Even setup under exactly the same experimental parameters, large signal fluctuations exist for multiple measurements of the same sample due to uncontrollable changes of the laser energy, delay time, the sampling gate width, and distance from the sample surface to the lens. Such fluctuations have a great impact on LIBS measurement precision and accuracy.

Normalization of each recorded spectrum over the whole spectrum area can reduce the signal fluctuations caused by ablation mass variation and the matrix effect [10, 25]. The spectral intensity after normalization can more accurately reflect the element mass concentration information. As a result, normalization by the whole spectrum area has become a widely accepted approach to reduce signal uncertainties and improve measurement precision and accuracy.

In the application for brass samples, the pre-processed data of multiple measurements were normalized with whole spectrum area, the averaged signal is taken as the spectral line intensity for calibration and prediction plots. In selecting the characteristic lines to establish the baseline, we found that the results of atomic line at 427.511nm and the ionic line at 221.027nm were the best two judging from $R^2$, RMSEP and RSD. Since the ionic line is more insensitive self-absorption effect, although the results of the atomic line at 427.511nm was even a little better than the ionic line at 221.027, the ionic line was chosen to set up the baseline.

The calibration curve and the prediction results of the traditional uni-variate model with whole area normalization and without any normalization were shown in Fig. 1. In the figure, the $R^2$ of the calibration curve after normalization and the RMSEP were 0.95 and 3.28%. Without normalization, the $R^2$ and RMSEP were 0.93 and 4.11%, respectively. As this example shows, normalization with the whole spectrum area improves the measurement accuracy. Comparing Fig. 1 (b) with Fig. 1 (a), it can also be found that the error bar is reduced after normalization. The results further demonstrate that normalization with the spectrum area can improve LIBS measurement precision. The RSD of the spectral line intensity for before and after the normalization with whole spectrum area was shown in Fig.3, and the results were consistent.

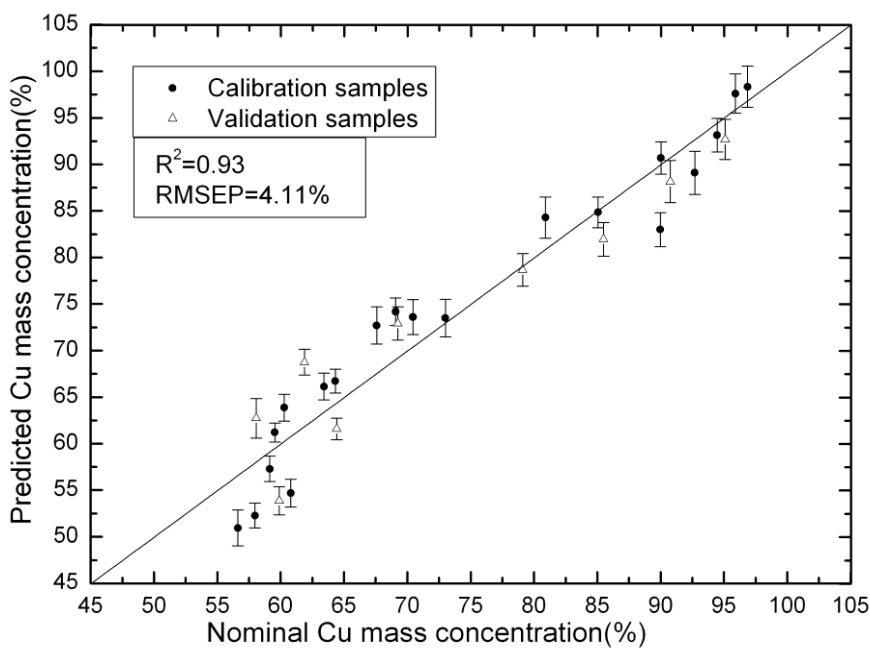

(a)

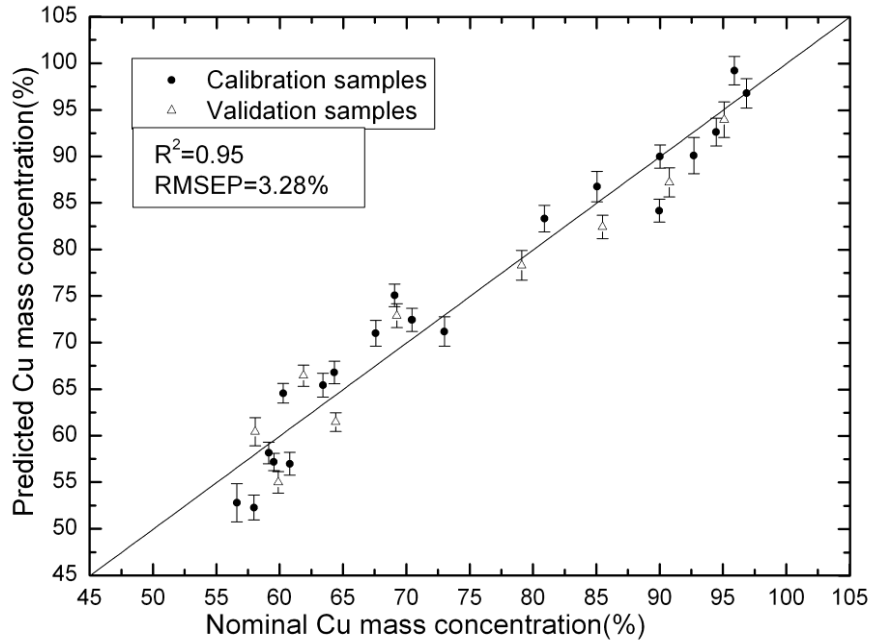

(b)

Figure 1. Calibration and prediction plots for traditional univarite model (Cu(II) at 221.027 nm) (a) without normalization; (b) normalization with the whole spectrum area.

Compared to the traditional uni-variate model without normalization, the model with whole spectrum normalization method also decreased the maximum relative error, as seen in Fig.5. Taking brass alloy sample ZBY927 for example, the maximum relative error decreased from 42.25% to 24.63% after normalization.

Overall, normalization by the whole spectrum area is a commonly used and effective method for reducing signal fluctuations and is therefore taken here as a baseline for comparison to clearly show the improvement of the present method.

## 4.2 Spectrum standardization model

### 4.2.1 Model construction

In order to convert the spectral line intensity to the intensity under standard plasma temperature and standard degree of ionization, it is necessary to calculate the plasma temperature and electron density (or degree of ionization) for each measurement and determine the standard temperature and standard degree of ionization.

To solve for the plasma temperature, the spectral lines who are not resonant lines and clearly separate from other adjacent spectral lines were selected to reduce influences of self-absorption and spectral line interference on the temperature calculation. Furthermore, excited energy of the upper level for the spectral lines should be made in a wide range to reduce calculation error for temperature [27]. According to the above principles, we selected seven characteristic spectral lines of the Cu atom (261.837, 282.437, 296.116, 427.511, 570.024, 578.213, and 793.312 nm) to construct a Boltzmann plane and calculate the plasma temperature. The results showed that the coefficient of determination ($R^2$) of Boltzmann plot ranged from 0.87 to 0.94. For a total of 1,015 measurements for all 29 samples, the plasma temperatures were between 8,600 K and 9,600 K, and the average temperature was 8,987.1 K. Therefore, an approximate mean, 9,000 K, was taken as the standard plasma temperature.

The electron number density was computed first in order to obtain a degree of ionization of elemental Cu. The Hα (656.27 nm) line (almost without self-absorption [31, 32]) was applied to compute electron number density. First, the full width at half maximum (FWHM) of the spectral line Hα (656.27nm) was utilized after Lorentz curve fitting, and then the electron density was calculated using the following formula [27, 29],

$$n_e(\text{cm}^{-3}) = C(T, n_e)(\Delta\lambda_{1/2})^{3/2} \tag{13}$$

where $C(T, n_e)$, the weak function of temperature and the electron number density, can be obtained from previous literature [27]. $\Delta\lambda_{1/2}$ is the FWHM of spectral line Hα (656.27 nm). The calculated electron number density ranged from $7 \times 10^{16}/\text{cm}^3$ to $1 \times 10^{17}/\text{cm}^3$ and it showed that the McWhirter criterion was satisfied in the experiments [29]. Using Eq. 2, the degree of ionization was obtained. Its range was from 0.62 to 2.14. The mean value of the degree of ionization, 1.05, was taken as standard degree of ionization.

In order to further reduce the uncertainty resulting from varying ablation mass, the sum of multi-spectral line were needed. In the application, the Cu spectral lines, which clearly separate from other adjacent spectral lines, were selected as listed in Table 2. The 32 Cu atomic or ionic spectral line intensities were converted to the intensity under standard plasma temperature and standard degree of ionization state using Eq. 4 or 5 to reduce or eliminate spectral line intensity variations caused by fluctuations of the two plasma physical parameters. The sum of intensities for those 32 spectral lines at standard temperature and degree of ionization was taken as $I_T(T_0, r_0)$ to represent for the signal that is proportional to the total number density of the specific element particles in ideal conditions.

Table 2. The characteristic spectral lines of Cu chosen to calculate $I_T(T_0, r_0)$.

| No. of spectral line | Atom/ion | Wavelength (nm) |
|---|---|---|
| 1 | Cu(I) | 261.837 |
| 2 | Cu(I) | 282.437 |
| 3 | Cu(I) | 296.116 |
| 4 | Cu(I) | 427.511 |

| 5 | Cu(I) | 522.007 |
| 6 | Cu(I) | 570.024 |
| 7 | Cu(I) | 578.213 |
| 8 | Cu(I) | 793.312 |
| 9 | Cu(I) | 809.263 |
| 10 | Cu(II) | 201.69 |
| 11 | Cu(II) | 202.549 |
| 12 | Cu(II) | 204.38 |
| 13 | Cu(II) | 206.242 |
| 14 | Cu(II) | 208.792 |
| 15 | Cu(II) | 210.039 |
| 16 | Cu(II) | 216.51 |
| 17 | Cu(II) | 216.991 |
| 18 | Cu(II) | 221.027 |
| 19 | Cu(II) | 224.7 |
| 20 | Cu(II) | 226.379 |
| 21 | Cu(II) | 227.626 |
| 22 | Cu(II) | 229.437 |
| 23 | Cu(II) | 236.989 |
| 24 | Cu(II) | 239.269 |
| 25 | Cu(II) | 240.012 |
| 26 | Cu(II) | 248.965 |
| 27 | Cu(II) | 250.627 |
| 28 | Cu(II) | 254.481 |
| 29 | Cu(II) | 330.787 |
| 30 | Cu(II) | 334.372 |
| 31 | Cu(II) | 589.046 |
| 32 | Cu(II) | 766.465 |

In detail, each of the spectral line intensity of each measurement was converted to the intensity $I(T_0, r_0)$ at standard plasma temperature (9000K) and standard degree of ionization (1.05) for all 29 brass samples using Eq. 4 or 5. Then, the sum of 32 spectral lines' intensity $I_T(T_0, r_0) = \sum_{i=1}^{32} I(T_0, r_0)$ was calculated for each measurement for all 29 samples. For each spectral line, the average value of $I(T_0, r_0)$ for each sample, were regarded as the intensity $I(n_{s0}, T_0, r_0)$ in the standard plasma state with standard plasma temperature, degree of ionization, and total number density of the measured element for this sample. Using curve fitting technology, $A_1$, $A_2$, $A_3$ were derived using Eq. 11, with all the measurements of calibration samples, therefore, the spectrum standardization model was built up. The value of $k_2$ was therefore calculated according to Eq. 10 as well as the coefficients $a_1$, $a_2$, and $b$ using the relationship with $A_1$, $A_2$, and $A_3$. In this manner, the calibration model was established. Table 3 lists the parameters determined for the spectral line Cu (II) at 221.027 nm.

Table 3. List of model parameters

| Parameter | $a_1$ | $a_2$ | $b$ | $k_2$ | $A_1$ | $A_2$ | $A_3$ |
|---|---|---|---|---|---|---|---|
| Value | 0.00789 | 0.0345 | -0.00667 | 0.0074 | 0.938 | -0.0324 | -0.000414 |

In addition, from the values listed, $A_1$ is much larger than $A_2$, which is consistent with the fact that the sum of 32 line of Cu serves as a signal to compensate for the fluctuation instead of a signal for concentration determination.

**4.2.2 Uncertainty reduction**

The standardization model established according to Eq. 11 considered the influence of varying temperature, degree of ionization and total number density of Cu atoms and ions on the intensity uncertainty with the aim of reducing the value.

For spectral line Cu(II) at 221.027 nm, the RSD of the intensity normalized through the whole spectrum area was reduced from the original 11.76% to 10.52% for sample ZBY901. Instead, using the present approach, the RSD decreased to 4.89% as shown in Fig.2. As shown in Fig. 3, the RSD of intensity normalized by spectrum standardization approaches for the 29 different samples (the average value of RSD was 5.29%) is significantly smaller than the RSD of intensity normalized by the whole spectrum area (with an average value 8.61% for RSD) and the RSD of the original intensity.

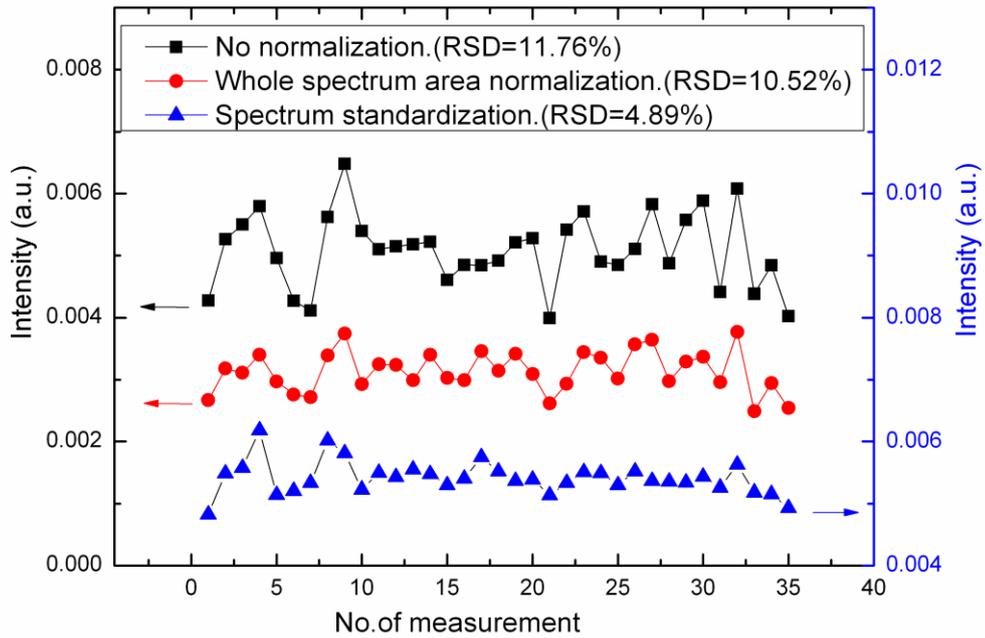

Figure 2. Pulse-to-pulse spectra fluctuations of the spectral line (Cu(II) at 221.027 nm) intensity for brass alloy sample ZBY901.

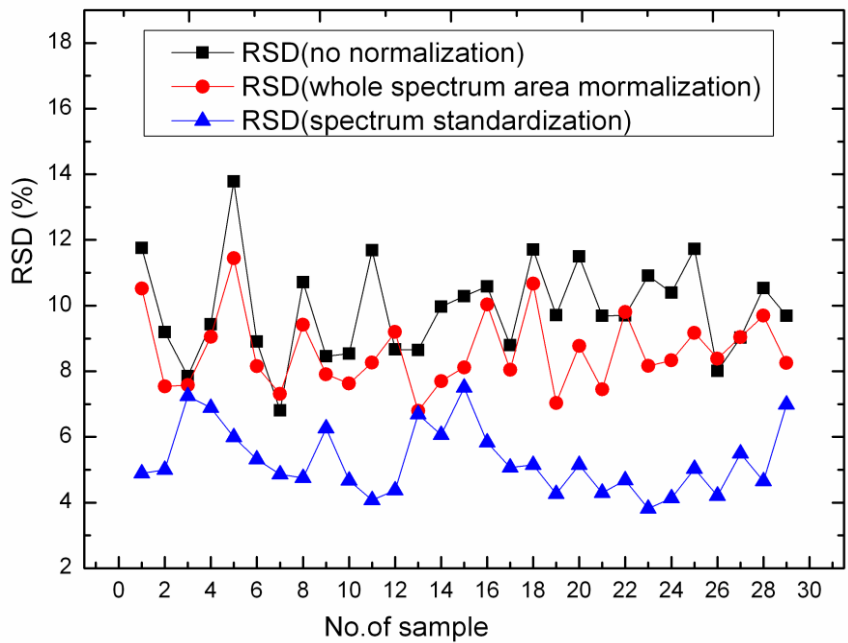

Figure 3. RSD values of the spectral line (Cu(II) at 221.027 nm) intensity normalized by different approaches for all 29 brass alloy samples.

Improvement in measurement precision can also be seen through a comparison with Fig. 1. (b), the error bars of the spectrum standardization model shown in Fig.

4(from 0.45 to 1.25, with an average value 0.68) were much smaller than the results of the whole spectrum area method model (from 0.93 to 2.05, with an average value 1.38). Seen from the results, it is apparent that standardization with plasma physical parameters can better reduce the signal fluctuations and improve the measurement precision than the conventional normalization with the whole spectrum area.

In addition, looking deep into the proposed normalization method, normalizing the spectral line intensity by plasma temperature and degree of ionization did not always successfully reduce the uncertainty for all spectral lines or one specific line for all samples. There may be two reasons for the results. First, the fluctuations of ablation mass may be the main source of the measurement uncertainty and the fluctuation of ablation mass may make the effect of RSD reduction due to plasma temperature and degree of ionization invisible. Secondly, the error and uncertainty in calculating the plasma temperature and degree of ionization will also weaken the effects of normalizing with temperature and degree of ionization.

After spectrum standardization, the remaining uncertainty of spectral line intensity could result from a combination plasma morphology diversity, inter-element interference effects, plasma parameters calculation uncertainties, and random noises.

### 4.2.3 Accuracy improvements

In this section, results of calibration and prediction for spectrum standardization model are exhibited.

For the spectral line Cu(II) at 221.027 nm, the $R^2$ of the calibration curve and the RMSEP were 0.98 and 2.72%, respectively (as shown in Fig. 4) for the spectrum standardization approach, which proved better than the results from the whole spectrum area standardized method ($R^2$=0.95, RMSEP=3.28%). These results illustrated the superiority of the present method in improving the prediction accuracy.

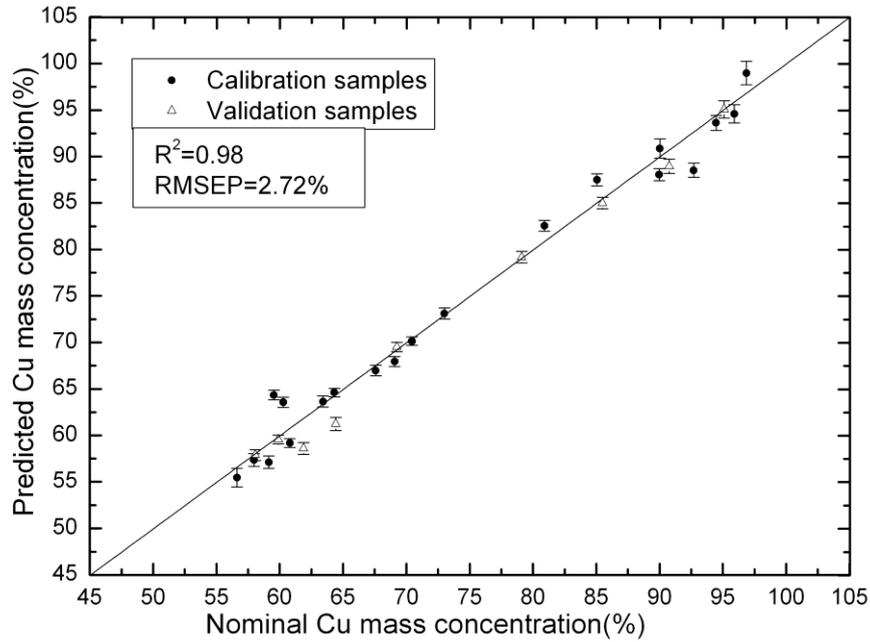

Figure 4. Calibration and prediction plots for spectrum standardization model (Cu(II) at 221.027 nm).

The maximum relative error is another index to evaluate the prediction accuracy, which was calculated by the following formula for each sample,

$$\text{maximum relative error} = \max\left(\left|\frac{C_{nomi} - C_{pre,i}}{C_{nomi}}\right| \times 100\%\right) \quad (14)$$

where $C_{nomi}$ is nominal elemental concentration for one sample, $C_{pre,i}$ is predicted concentration for the No.$i$ measurement, where $i$ ranges from 1 to 35 for each sample. As shown in Fig.5, the maximum relative error of model prediction for Cu mass concentration in the present model (the average value of 29 samples' maximum relative error was 16.97%) were much smaller than in the whole spectrum area normalization model (with an average value of 27.19%), further proving the present model's advantage in improving prediction precision.

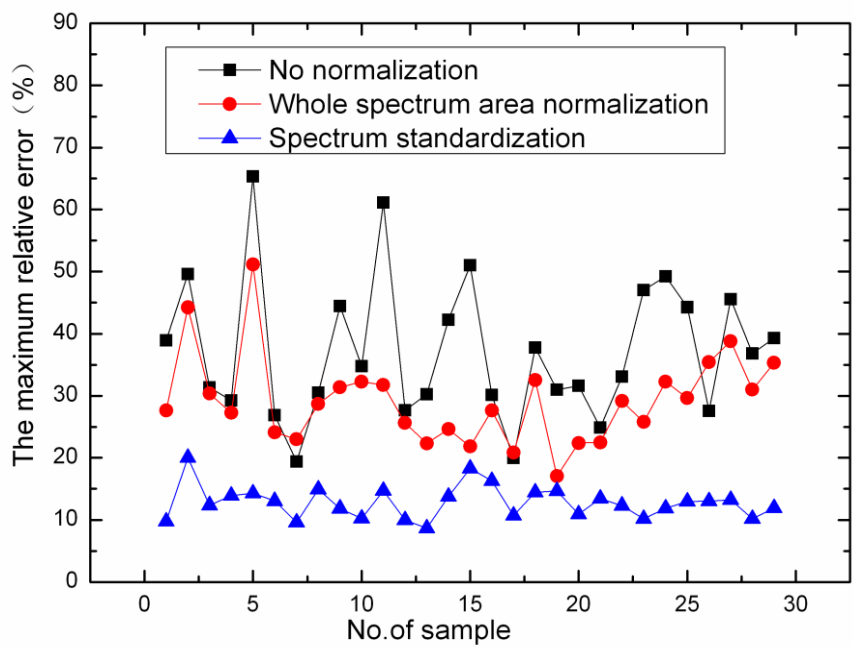

Figure 5.The maximum error of predicted Cu mass concentration by the spectral line Cu(II) at 221.027 nm for different models.

In addition, we have tried all the following characteristic lines including both atomic and ionic lines (204.38nm, 221.027nm, 227.626nm, 236.989nm, 250.627nm, 254.481nm, 330.787nm, 427.511nm, 522.007nm, 793.312nm, 809.263nm), which have $R^2$ higher than 0.8 in the uni-variate model established with original spectral data, with the spectrum standardization method, and results shows that all lines except Cu(I) line 427.511nm, the standardized method yields both $R^2$ closer to unity and smaller RMSEP than normalization with whole spectrum area, which in turn, better than without normalization. For spectral line 427.511nm, the results for the three methods are very close with the whole area normalization performing best.

Furthermore, the results of the present model for line at 221.027nm were compared with the results of normalization with whole spectrum area for atomic line at 427.511nm since it yielded the better baseline model results, and results showed that the present result showed a clear improvement for all indexes such as RSD, $R^2$, standard error, RMSEP, and maximum relative error.

## 5 Conclusion

The present work proposed a so called "spectrum standardization" method to utilized the plasma character parameter to compensate for the fluctuations of characteristic line intensity due to variation in plasma temperature, degree of ionization, and total number density of the measured element. In application for 29 brass samples, the proposed standardization method shows its potential to greatly improve the measurement precision and accuracy over generally applied normalization with whole spectrum area method.

It has been notice that the processes for the standardization were very complicated and cost of time and energy. Research that can simplify the standardization process is now undertaking and the work will be reported in a companion paper. Besides, the method favor ionic lines and requires sum of multi-lines to compensate for the fluctuations of total number density, there also needs more research for cases that only a few characteristic lines available. For example, there are only a few atomic carbon lines with clear spectral profile for C element measurement in coal, the application of this method may be limited and there is more research work needed.

# Acknowledgement

The authors acknowledge the financial support from the Chinese governmental "973" project (NO. 2010CB227006) and "863" project (NO. 20091860346).